# The general traits of inelastic electron scattering by the adsorbed system

A.R. Cholach[*] and V.M. Tapilin

*Boreskov Institute of Catalysis, Prospekt Akademika Lavrentieva, 5*
*630090 Novosibirsk, Russian Federation*

**Abstract**

Inelastic electron scattering by the adsorbate covered Pt(100) single crystal surface is studied by Disappearance Potential Spectroscopy and density of states (DOS) calculations. Two peculiar channels of elastic electron consumption are highlighted, both related to the substrate core level excitation coupled separately with two particular electron transitions. The first channel affects the adsorbed layer and enables to reveal the valence state structure of the adsorbed species as well as the substrate DOS. The second one includes the multiple plasmon oscillations. The proposed mechanism of electron transitions assumes that one-dimensional DOS at the vacuum level is an additional spot for location of excited electrons, along with vacant DOS at the Fermi level. Observed phenomena are supposed to be a general regularity of electron-solid interaction and a useful tool for fingerprinting the adsorbed layer at molecular level.

[*]Corresponding author. Tel.: +7 383 326 95 40; fax: +7 383 330 80 56;
E-mail address: *cholach@catalysis.ru*

# 1. Introduction

Disappearance Potential Spectroscopy (DAPS) is based on the threshold core level excitation of target atoms by the primary electron beam of time-based energy $E_p$ [1]. As a result of energy loss, a part of incident electrons disappears from elastic current $I(E_p)$ and provides a sharp drop in the $dI(E_p)/dE_p = f(E_p)$ dependence being recorded experimentally. In the course of excitation operative, the core and primary electrons occupy available vacant states just above the Fermi level $E_F$, and so the self-convolution of vacant states density is in charge of the spectral structure. The spectral feature is accordingly localized at the energy point equal to difference between the core level and the vacant state energy. The high surface sensitivity ~1-3 monolayer defined by half the electron mean free path in a solid makes DAPS a promising tool in adsorption and heterogeneous catalysis investigations. However, this technique is not widely adopted so far mainly due to the following reasons.

According to DAPS principle the spectral structure depends not only on a surface content, but greatly on a given density of vacant states structure, and so DAPS looses as analytical techniques in comparison with Auger-electron (AES) and X-ray Photoelectron Spectroscopy (XPS). The peculiarities of the vacant state structure is generally the main outcome of conventional DAPS application [1]. Furthermore, an appreciable density of vacant states at $E_F$ is a precondition for obtaining the clearly defined spectrum. It is not the case for nearly filled outer shell of the Platinum group metals which are of great practical importance as active components of a good deal of industrial catalysts [2]. In particular, there is the only paper with a very few comments related to the $Pt3d_{5/2}$ DAPS study by the primary electron beam of above 2 keV energy [3]. We have established earlier a perfect agreement between calculated and experimental DAPS spectra around the $Pt4d_{5/2}$ core level obtained for the hydrogen and oxygen covered Pt(100)-*(1x1)* single crystal surface [4], but this result is mainly of theoretical importance since the valence rather than the vacant states are responsible for adsorption, catalytic and other special properties of a solid.

A shake-off process is quite a common occurrence that is often observed by means of AES and XPS [5,6]. A set of regular spectral features similar to shake-off satellites of isolated molecules [7] was revealed in $Pt4d_{5/2}$ DAPS spectra after formation of different adsorbed layers on the Pt(100)-*(1x1)* single crystal surface [8]. The number and locations of these features above the threshold are well correlated with number and ionization potentials of corresponding valence states of the adsorbed particles determined for analogous systems by Ultraviolet Photoelectron Spectroscopy (UPS). Thus a single *H1s* or *O2p* satellite was detected after $H_2$ or $O_2$ adsorption; three satellites related to ionization of *1π*, *5σ*, *4σ* and *2π*, *1π+5σ*, *4σ* states were revealed after CO and NO adsorption, respectively. This phenomenon was regarded as an additional channel of inelastic electron scattering, namely the coupled (conjugate) electron excitation (CEE) which includes the threshold core level excitation of substrate atom and ionization of the valence state of the adsorbed species. Besides that, a multiple bulk and surface Pt plasmon excitations are believed to be also coupled with the core level excitation [9].

The previous paper [8] is just a sequence of very promising results needed to be confirmed and clearly understood. This work highlights the novel kind of electron transition affecting an occupied, substrate and adsorbate density of states (DOS) which has never been considered in a standard theory of DAPS. The proposed mechanism of such a transition assumes a pronounced DOS at the vacuum level to be an additional spot for location of excited electrons, along with the vacant DOS at $\underline{E}_F$. The present work includes reinterpretation of fine DAPS spectral structure in light of an advanced DOS calculations, partially reiterates previous data using an improved spectra processing, and shows new experimental results related to $H_2$ adsorption. The paper as a whole reveals the advanced possibilities of simple and accessible DAPS technique for investigation the adsorptive, catalytic and other special properties of solids at the molecular level.

## 2. Experimental and theoretical

Experiments were performed using the ultrahigh vacuum chamber with residual gas pressure of $< 2 \cdot 10^{-8}$ Pa. The chamber was equipped with low-energy electron diffraction (LEED), AES with retarding field analyzer, monopole mass-spectrometer and Ar+ ion gun. The DAPS technique was realized on the basis of 3-grid LEED optics. The central LEED electron gun with a tungsten filament was used to form the primary electron beam of ~ 1 mm size, ~ 1 µA current with normal incidence to the sample surface. The time-based accelerating potential was applied to W-filament. The outside LEED grids were grounded while the middle one was at the same negative potential as the electron gun filament, so that only quasi-elastic electrons scattered by the target could pass it to reach the LEED screen as a collector. The pass band cutoff 2-3 eV of the retarding grid energy analyzer was kept constant over a given set of experiments. The spectra were recorded as a first derivative of total collector current by the modulation of sample potential (~ 0.5 $V_{p-p}$; ~ 2 kHz). The average spectral resolution was ~ 1.0 eV, the peak position could be determined within ~ 0.1 eV precision.

The Pt(100) single crystal of 99.999% purity, ~ 1cm width and ~ 1 mm thickness was oriented within ~ 1°. The thermodynamically favorable Pt(100)-*hex* surface structure is known to reconstruct readily into *(1x1)* one under adsorption of certain molecules [10]. In order to keep stable the substrate structure all adsorption experiments were performed at room temperature on the AES-clean *(1x1)* surface which was obtained according to "NO-receipt" [11].

The DAPS spectra methodically accumulate an attendant diffraction features which are up to $10^3$ times more intensive than the true DAPS peaks for well-ordered single crystal surface. And so the difference spectra (the adsorbate covered surface minus the clean one) were considered where diffraction features suppress each other under subtraction. Of course, the Pt spectral contribution also becomes partially self-suppressed; therefore those features are revealed which were changed under adsorption taking part in the adsorbate-surface bond formation. The difference spectra for the CO/Pt(100) system were still complicated by the strong diffraction peaks originated from the well-ordered adsorbed layer. In order to eliminate this background a smooth Gaussian fit of each spectrum was subtracted from the difference one. Above procedures are commonly used in data processing to find out the fine spectral structure without disturbance [12]. The DAPS measurements were performed in the neighborhood of Pt4d core level excitation to minimize the diffraction background, a detailed description of sample treatment can be found elsewhere [2,8]. The reference $E_F$ point in all spectra corresponds to the incident electron energy of 314.8±0.5 eV that is close to reference value of the Pt4$d_{5/2}$ core level energy 314.6 eV [12] with due regard for the filament work function [1,8].

The calculations of bulk Pt DOS were performed by the Quantum Espresso package [13] for experimental lattice parameters. Core electrons were treated with the Perdew-Burke-Ernzerhof ultrasoft pseudopotential [14], the plane waves with $E_{cut}$ = 40 Ry were used for basis set of the valence electron wave function. A numerical integration in the reciprocal space was performed within a *12x12x12* grid. The present technique permits to calculate precisely the vacant state DOS and to remediate deficiencies of earlier calculations [8] suffered from the poor basis set. The primary electron is described by the plane wave with vector component $\vec{k}_\parallel = 0$ parallel to the surface.

## 3. Results and discussion

*3.1. The substrate DOS probing by DAPS*

The threshold Pt4d core level excitation is the main channel of the primary electron energy consumption traced by DAPS in which two electrons, the primary $e_p^-$ and the core level electron take part:

$$e_p^- + 4d_0^{10} + a_{1,2}^0 \rightarrow 4d_{5/2}^9 + a_{1,2}^2, \qquad (1)$$

where *$a_{1,2}$* stands for the Pt vacant states.

It is commonly accepted that no information on the occupied part of valence bands can be obtained from DAPS spectra. However, numerous regular CEE satellites gives evidence that the main channel (1) is coupled with ionization of the adsorbate valence band; therefore a three-electron process can also join the primary electron energy consumption (1). In this process, in addition to two operating electrons excited to the vacant $\underline{a}$-states just above $\underline{E}_F$, a valence electron at state $\underline{V}$ of the adsorbed species moves to vacuum level and thus leaves the system as $e^-_{em}$. Therefore, the electron transitions (1) should be corrected as:

$$e^-_p + 4d_0^{10} + V^2 + a^0_{1,2} \to 4d^9_{5/2} + V^1 + a^2_{1,2} + e^-_{em} \qquad (2)$$

In the general case of transition (2) three electrons occupy some vacant states with energies $\underline{\varepsilon}_1$, $\underline{\varepsilon}_2$ and $\underline{\varepsilon}_3$, and one of these electrons is excited at that from the valence state with energy $\underline{\varepsilon}_V$. And so the probability $\underline{W(E)}$ of process (2) is determined by the DOS self-convolution:

$$W(E) = \int_0^E d\varepsilon_1 \sigma(\varepsilon_1) \int_0^{\varepsilon_1} d\varepsilon_2 \sigma(\varepsilon_2) \int_0^{\varepsilon_2} d\varepsilon_3 \sigma(\varepsilon_3) \sigma(E - \varepsilon_1 - \varepsilon_2 - \varepsilon_3 + \varepsilon_V), \qquad (3)$$

where $\sigma$ is the valence band DOS and $E$ is energy excess above the threshold energy.

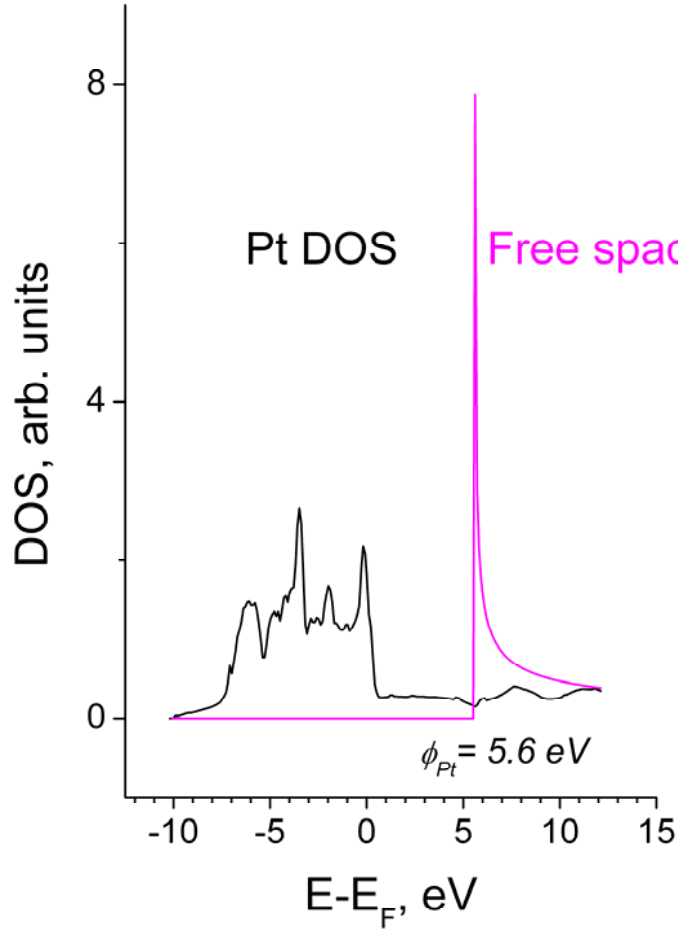

**Fig. 1** The DOS of bulk Platinum (black) and of free space (magenta) relative the Fermi level.

Strictly speaking one should consider the local DOS which defines the bulk-to-vacuum DOS transformation. For qualitative understanding of the spectral structure we have considered only extreme cases, namely the bulk Pt DOS and the vacuum space shown in Fig. 1.

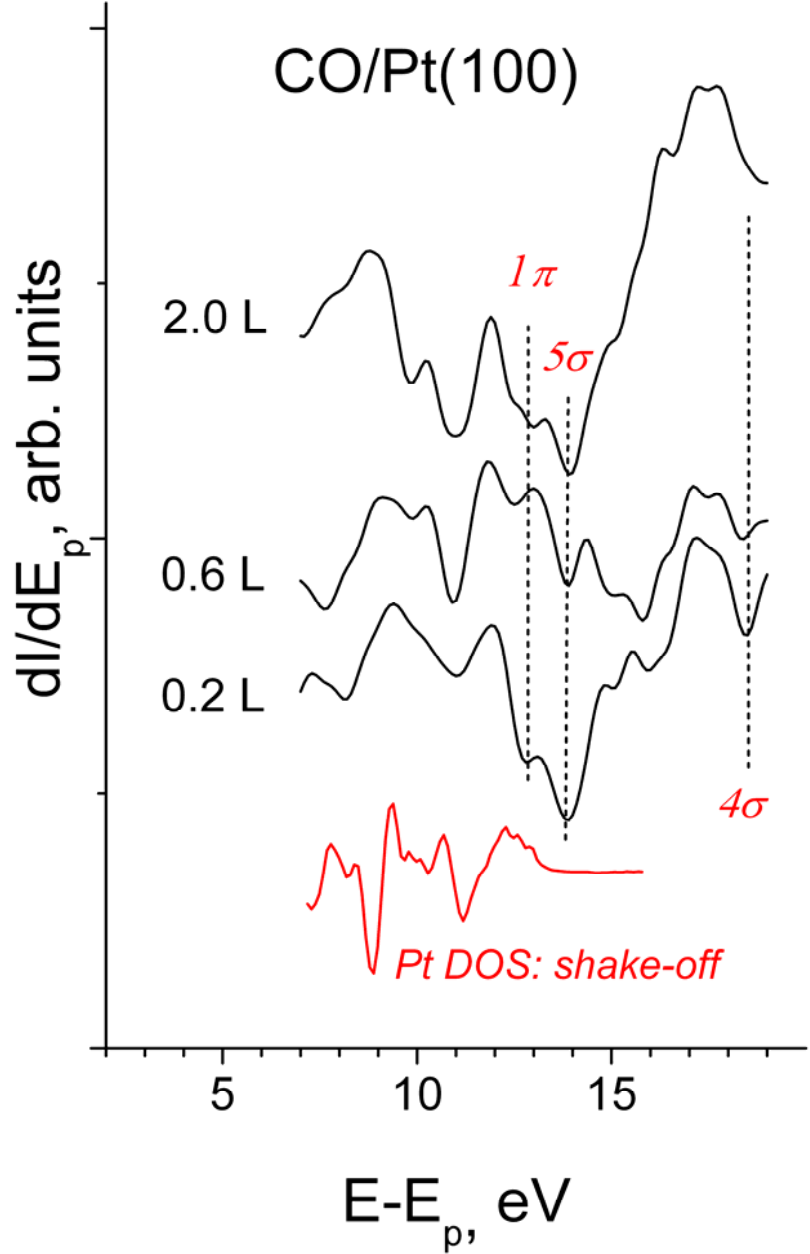

**Fig. 2** The DAPS spectra obtained after various exposure of clean Pt(100) surface to CO; red curve: shake-off satellite structure expected from data of Fig. 1.

The scattering of electron waves with wave vectors $\vec{k}_1$ and $\vec{k}_2$ in states with wave vectors $\vec{k}_1 \pm \vec{q}$ and $\vec{k}_2 \mp \vec{q}$ is governed by the Fourier transform of Coulomb interaction $1/|q|^2$, therefore the scattering with small changes of wave vectors is preferable. It means that the number of vacant states to which the scattering is possible is proportional not to the total density of states, but rather to its part near the initial wave vectors. That is the reason to use one-dimensional DOS $\sim 1/|\nabla_{\vec{k}} E(\vec{k})|$ for the vacant states where $E(\vec{k})$ is the band energy, instead of the total DOS obtained by integration of $1/|\nabla_{\vec{k}} E(\vec{k})|$ over the constant energy surface in the wave vector space. In the vacuum region for a free electron the total DOS $\sim E^{1/2}$, while the 1-D DOS $\sim E^{-1/2}$ [15] as shown in Fig. 1. The unoccupied part of the bulk Pt DOS is significant only right above the Fermi energy as in the case of well-ordered Pt(111) surface [16]. The DOS is small and nearly constant at larger energy, and so the main transition of core electron will occur only in the narrow energy region around $\underline{E_F}$. On the other hand, the vacuum states shown in Fig. 1 exhibit a strong maximum (infinite theoretically) at the work function of clean Pt (100)-*(1x1)* surface $\varphi = $ 5.6 eV [17,18]. A generally accepted theory of the threshold core level excitation does not consider these states as a possible spot for location of excited electrons [1], but our previous experimental observations [8] reaffirmed by the present ones give evidence that it can be the case. Therefore the DOS above $\underline{E_F}$ in Fig. 1 can be approximately represented by two discrete levels at $E_F$ and $E_F + \varphi$ that permits to reduce expression (3) to the following one:

$$W(E) \approx \sigma^3(E_F)\sigma(-E)_{E>0} + \sigma^2(E_F)\sigma(E_F+\varphi)\sigma(\varphi-E)_{E>\varphi} \qquad (4)$$

Fig. 2 shows the difference DAPS spectra obtained after various exposure of clean Pt(100) surface to CO. In addition to expected *1π*, *5σ*, *4σ* CEE satellites there is a pronounced spectral structure within the range of 8-12 eV. It should be noted that numerous theoretical calculations and UPS measurements definitely relate the similar spectral features to Pt DOS [19-21]. If the Pt DOS also takes place in data of Fig. 2 then outgoing electron should overcome a sum of the valence state energy relative $\underline{E_F}$ and the work function barrier. Thus at the total energy point one should observe a spectral feature in experimental *dI(E_p)/dE_p* dependence corresponding to the shake-off satellite. The red curve in Fig. 2 was constructed from data of Fig. 1 in the following way. Accounting for aforementioned details of spectra recording, the Pt DOS below $\underline{E_F}$ was reversed, differentiated and shifted to the right by the Pt work function according to the second term in relation (4). Therefore, the red curve simulates the expected spectral structure assuming a direct electron transition from the Pt valence band to a narrow localized state at the vacuum level. Fig. 2 displays a perfect correspondence between simulated and experimental spectral features in the range of 8-12 eV.

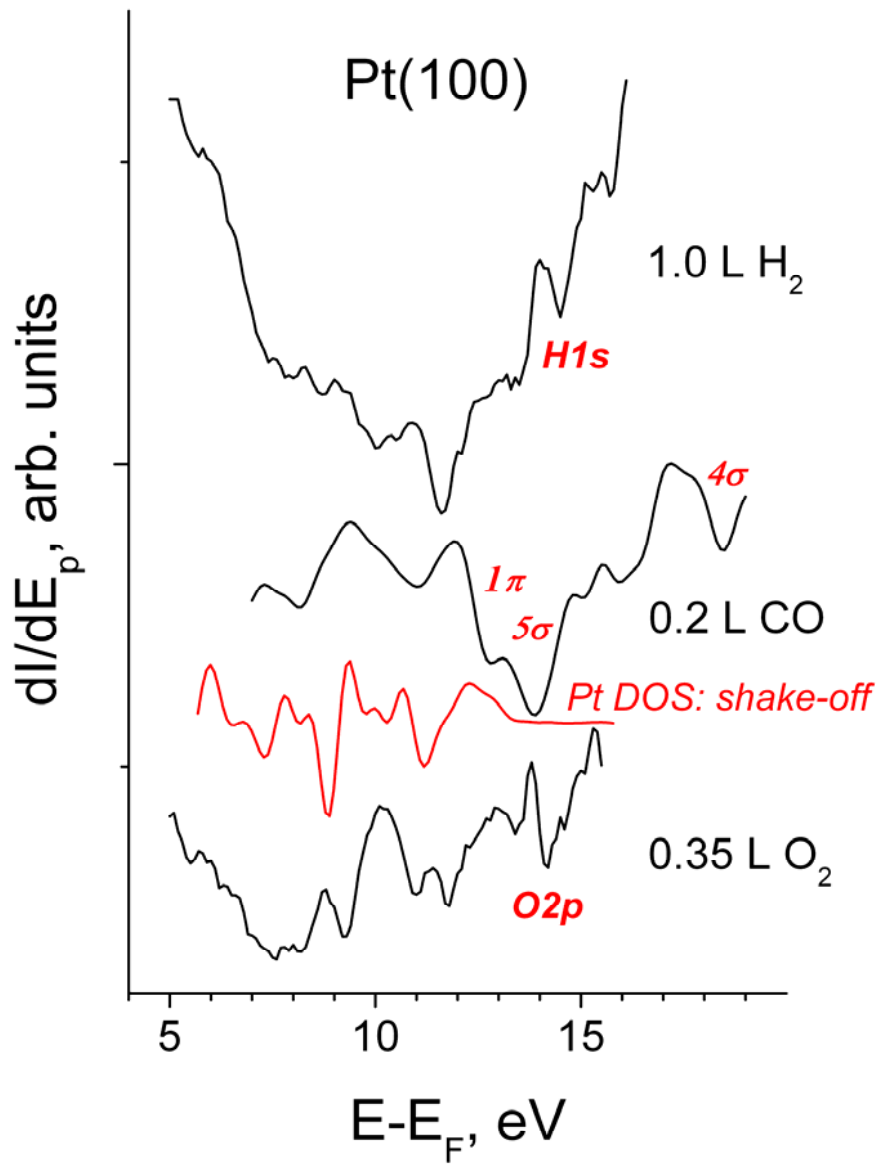

**Fig. 3**  The DAPS spectra obtained after shown exposure of clean Pt(100) surface to a given adsorbate; red curve: shake-off satellite structure expected from data of Fig. 1.

Fig. 3 exhibits a set of referred CEE satellites in DAPS spectra related to different adsorbed layers. In addition, the similar structure at the 8-12 eV energy range is clearly seen in each spectrum as well as in the red curve copied from Fig. 2. Therefore Fig. 2 and Fig. 3 give a direct evidence for that the valence state $\underline{V}$ in equation (2) may belong to the substrate atom as well as to the adsorbed particle.

*3.2. The generality of CEE effect*

Fig. 4 shows a set of raw analogous DAPS spectra obtained after exposure of clean Pt(100) surface to $O_2$. The relevant spectral features are summarized in Table 1. The Pt DOS in Fig. 1 showed no fine vacant states structure close to the Fermi level. In order to assign a set of distinct peaks at 2-7 eV in Fig. 4 we assume that a narrow vacant state just above $E_F$ alike that at the vacuum level can be also a spot for shake-up transitions of the Pt valence band electrons. The blue curve in Fig. 4 simulates this kind of electron excitation in the same way as above, but without the work function shift according to the first term in relation (4). An acceptable correspondence between experimental and simulated spectral structure testifies the possibility of such a transition.

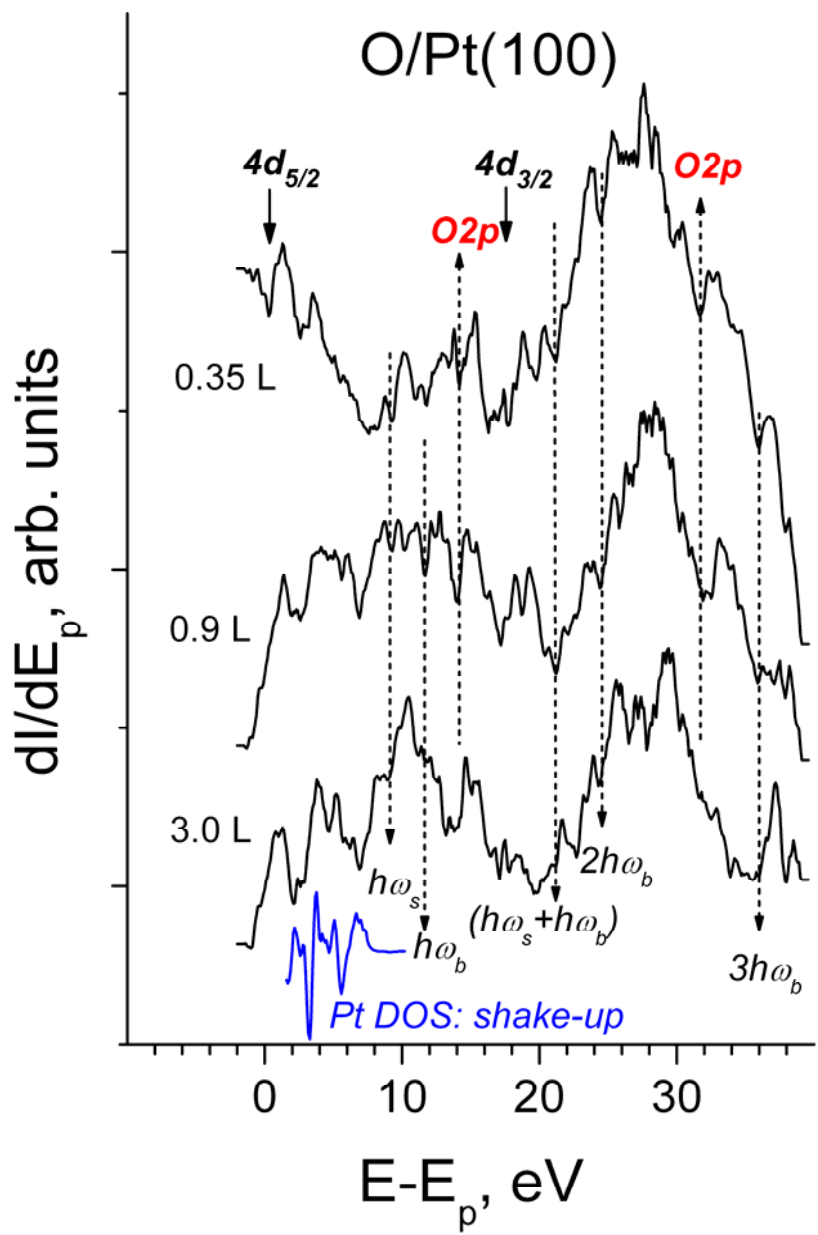

**Fig. 4**  The raw difference DAPS spectra obtained after indicated exposures of clean Pt(100) surface to $O_2$; blue curve: shake-up satellite structure expected from data of Fig. 1.

**Table 1**

Relevant spectral features in the extended DAPS spectra of Figs. 2-5.

| E-$E_F$, eV | Peak assignment | Comment |
|---|---|---|
| 0 | *Pt4d$_{5/2}$* | The reference point (Fermi level) |
| 2-7 | *shake-up* | Occupied Pt5d-band satellites; Figs. 4, 5 |
| 8-12 | *shake-off* | Occupied Pt5d-band satellites; Figs. 2, 3, 5 |
| 17.2 | *Pt4d$_{3/2}$* | *Pt4d$_{3/2}$* core level (*Pt4d* spin-orbit splitting); Fig. 4 |
| 14.0 | *O2p* | CEE of the *O$_{ad}$* species, relative *Pt4d$_{5/2}$*; Fig. 4 |
| 31.5 | *O2p* | CEE of the *O$_{ad}$* species, relative *Pt4d$_{3/2}$*; Fig. 4 |
| 9.2 | *ℏω$_s$* | Relative *Pt4d$_{5/2}$*; Fig. 4 |
| 11.8 | *ℏω$_b$* | Relative *Pt4d$_{5/2}$*; Fig. 4 |
| 21.2 | *ℏω$_s$+ℏω$_b$* | Relative *Pt4d$_{5/2}$*; Fig. 4 |
| 24.2 | *2ℏω$_b$* | Relative *Pt4d$_{5/2}$*; Fig. 4 |
| 36.0 | *3ℏω$_b$* | Relative *Pt4d$_{5/2}$*; Fig. 4 |

A pronounced peak in Fig. 4 at 14.0 eV stands for the CEE satellite related to ionization of the *O2p* state [22] coupled with the Pt4d$_{5/2}$ threshold excitation core level as discussed above [8]. Furthermore, a splitting of this feature at top exposure may be considered as the formation of new adsorbed state of oxygen atoms *O$_{ad}$*, e.g. like the bridge state or incorporated one [23,24]. In this way the peak at 31.5 eV is also *O2p* satellite, but coupled now with the Pt4d$_{3/2}$ core level excitation. As expected, the intensity of both *O2p* satellites increases on coverage and they are localized apart each other at characteristic spin-orbit Pt4d splitting [25]. According to electron theory of metals every core level can undergo a threshold excitation, and there are no reasons prohibiting the CEE process related to excitation of the other core level than the Pt4d one. The present observation in Fig. 4 corroborates this statement and implies that the other satellites alike *O2p* one, and therefore the valence state structure of the adsorbed species will be coupled with the threshold excitation of any substrate core level. The multiplicity and regularity of similar spectral satellites revealed from different adsorbed layers suggests that the CEE effect is believed to be a fundamental regularity of inelastic electron scattering in any adsorbed system.

In theory, CEE monitoring enables to determine a localization of the adsorbed species at multicomponent surface, because CEE satellites should accompany threshold excitation of only that surface atom which is chemically bound to a given particle. In turn, the reference core level energies of different elements are easily distinguishable. It should be noted that it is a high surface sensitivity of DAPS technique that provides a substantial spectral contribution of CEE satellites originated from the adsorbed layer.

*3.3. Plasmon excitation as another kind of the conjugate electron excitation*

Plasmon excitation is an effective channel of inelastic electron scattering. It is responsible for the most intensive peaks displayed for different surfaces by (Reflection) Electron Energy Loss Spectroscopy (R)EELS [26]. For instance, EELS spectra obtained for the clean Mo(110) [27], W(111), (112) [28] and Si(100) [29] single crystal surfaces reveal a set of 10-20 features with the base width ~ 5 eV related to multiple surface and bulk plasmon excitations. Interband transitions make difficult an identification of plasmon oscillations in EELS spectra. A transition with energy $E$ is usually assigned as a bulk plasmon excitation if multiple transitions divisible by $E$ and the surface plasmon transition $E/\sqrt{2}$ are also revealed [30].

A set of not intensive, but systematic narrow peaks are perceptible in spectra of Fig. 4 and listed in Table 1. The general regularities of these features consist in the following. First, in contrast to $O2p$ satellites the peak intensity decreases on oxygen coverage. Second, three of them are localized at invariable distance of 12.0±0.2 eV apart each other including the first satellite relative $E_F$. This behaviour is typical for the multiple bulk plasmon excitations resulted in the energy loss aliquot to plasmon frequency $\hbar\omega_b$ [27-29]. Then the feature at 9.2 eV corresponds to surface plasmon $\hbar\omega_s$ with an expected ratio of $\omega_b/\omega_s \sim \sqrt{2}$ [31]. The peak at 21.2 eV is similarly assigned to the multiple bulk and surface plasmon excitation of the total $\hbar\omega_s+\hbar\omega_b$ energy. The weak features in the range of 26-30 eV can be attributed to the coupled plasmon and $Pt4d_{3/2}$ core level excitation. It is important to note that the reference point of plasmon locations is the substrate Fermi energy which corresponds to threshold excitation of the $Pt4d_{5/2}$ core level. In other words, the coupled plasmon and core level excitation can be considered as another kind of CEE phenomenon discussed above.

Fig. 5 shows the difference DAPS spectra obtained after exposure of clean Pt(100) surface to $H_2$. The above consideration enables to assign properly all relevant features. The $H1s$ satellite is the same as in Fig. 3. The shake-off emission from the Pt valence band is responsible for the spectral structure at 8-12 eV similar to Fig. 2 and Fig. 3. The shake-up transition of the Pt valence band stands for the spectral features at 2-7 eV as in Fig. 4. The peak at ~ 13.3 eV probably results from the bulk plasmon excitation $\hbar\omega_b$, and its intensity decreases on $H_{ad}$ exposure as expected. Then the peak at ~ 10.0 eV at 0.5 L $H_2$ exposure corresponds to surface plasmon $\hbar\omega_s$, and it disappears at higher $H_2$ exposure. The decrease in plasmon features intensity on coverage in Fig. 4 and Fig. 5 is certainly due to the shielding effect of the $O_{ad}$ and $H_{ad}$ layer, respectively.

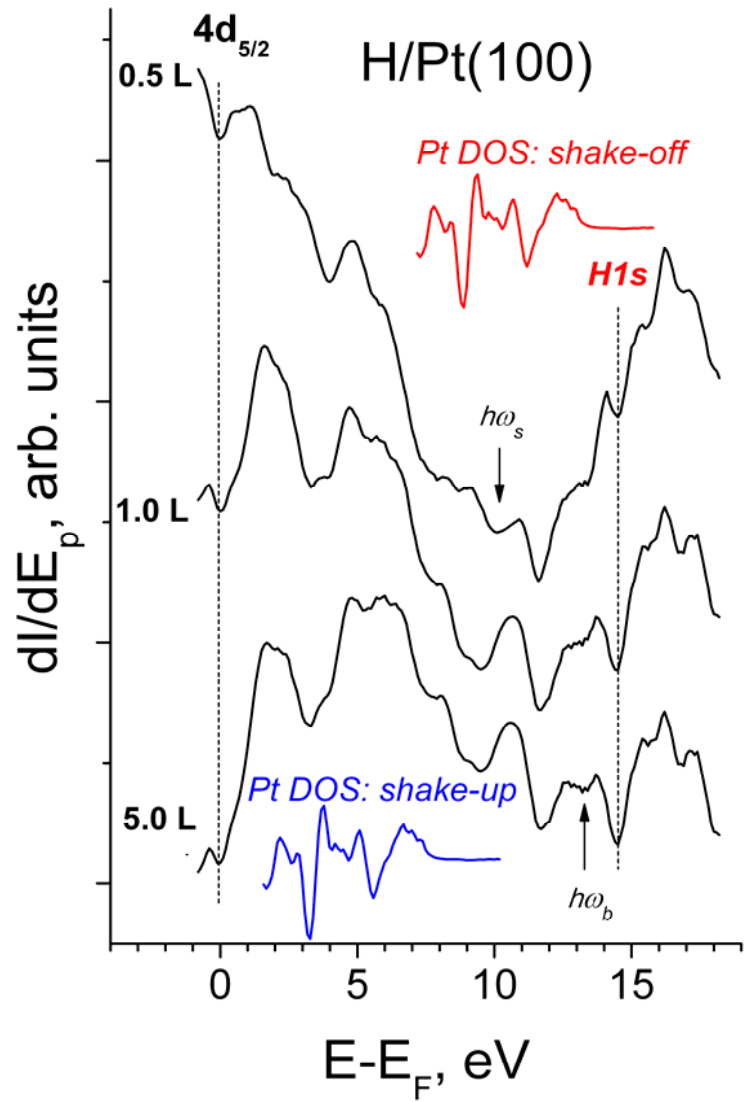

**Fig. 5**   The DAPS spectra obtained after indicated exposure of clean Pt(100) surface to $H_2$; blue and red curve: shake-up and shake-off satellite structure expected from data of Fig. 1, respectively.

The reliability of plasmon energy can be evaluated from Table 2. The present plasmon energy has to be refined in the following way. First, it should be shifted up by half the typical base width 1.5-2.5 eV of REELS peak for the proper comparison with reference data, because Fig. 4 and Fig. 5 represent DAPS spectra in a derivative mode. On the other hand, it should be shifted down by the double energy of vacant *a*-state ~ 1 eV where two operating electrons are finally localized according to the prime process (1). Taking into account these corrections the agreement between the reference data and the present value of $\hbar\omega_b$ = 12.0÷13.3 eV seems satisfactory. Theoretical estimation of the bulk plasmon excitation energy is performed by the free electron gas formula [30,37]:

$$\omega_b = \left(\frac{4\pi n e^2}{m}\right)^{1/2}, \tag{5}$$

where $\underline{n}$ and $\underline{m}$ stands for effective number of free electrons per Pt atom (supposed to be *s* and *p* electrons) and effective electron mass, respectively.

**Table 2**

The energy of surface $\hbar\omega_s$ and bulk $\hbar\omega_b$ plasmon excitation, and characteristic peak base width for different surfaces, eV

| Surface | Peak base width | $\hbar\omega_s$ | $\hbar\omega_b$ | Comment |
|---|---|---|---|---|
| Al-film | ~ 3 | 10.3 | 15.3 | EELS [32] |
| Al-film | ~ 5 | 10 | 15 | REELS [33] |
| Al(100) | ~ 5 | 10.5 | 14.9 | REELS [31] |
| Pt-foil | ~ 3 | | 15.2 | REELS [34] |
| Pt(100) | ~ 5 | | 13.9 | REELS [35] |
| | | | 13.5 | Theory [36] |
| O/Pt(100) | ~ 1 | 9.2 | <12.0> | DAPS, Fig. 4 |
| H/Pt(100) | | 10.0 | 13.3 | DAPS, Fig. 5 |
| Pt | 2 free electrons | 9.1 | 12.9 | Formula (5) |
| | 3 free electrons | 11.2 | 15.8 | |

Experimental plasmon energy is usually lower than followed from the free electron gas model [38]. Table 2 indicates the number of free electrons between 2 and 3 per Pt atom. The presently calculated electron configuration in Platinum crystal $5d^8 6s^2$ supports this statement if d-electrons are treated as localized.

Plasmon oscillations can be detected by means of AES and XPS [31]. In both cases a part of secondary electron energy aliquot to linear combination of $\underline{\hbar\omega}_s$ and $\underline{\hbar\omega}_b$ is consumed for plasmon excitations. Therefore spectral satellites are observed at corresponding positions at the lower-energy side of the main KLL Al(111), KLL Si(100) [39,40] peak or along the higher-energy side of the main C1s [40,41], Ge2s [42], Al2s and Al2p [33] peak. The physics and surface sensitivity of plasmon losses in XPS and EELS are quite similar, except the presence of electron–hole interaction and the lack of collimated beam in the former case which is qualified as a core-level loss spectroscopy [43]. In both cases plasmon oscillations come true due to energy loss by the primary- or photo-electron of arbitrary energy, and so the plasmon spectral features can be considered as shake-up satellites. In contrast to that, plasmon oscillations listed in Table 1 can be observed conditionally: in the presence of the core level excitation as a basic process, and if there is an excess of primary electron energy above the Pt4d threshold. That is why plasmon features are localized at higher energy region with respect to the Fermi level. Furthermore, the scattered electron providing plasmon oscillations in EELS, AES and XPS measurements is actually an external one. On the contrary, an electron participating in CEE process can be duly regarded as internal one since it has just taken part in the core level excitation and became indistinguishable among other electrons within the integrated, substrate and adosbate DOS. The difference in excitation mechanism enhances an efficiency of electron energy dissipation and results in a relatively high resolution of plasmon satellites in DAPS spectra. Thus, the typical peak base width in CEE processes is ~ 1 eV (Fig. 4, Fig. 5), whereas that of plasmon features observed by EELS and XPS is 5-10 eV and 20-30 eV, respectively [39-42]. According to Fig. 1, the low density of vacant states close to $E_F$ necessary for the leading process of the Pt4d$_{5/2}$ core level excitation results in low intensity of plasmon satellites in present DAPS spectra. The same reason is mainly responsible for the shortage of true DAPS peaks. Indeed, only a small shoulder in the range of 0 - 1 eV can be related to the prime process (1) in Fig. 4 and Fig. 5. Of course, the aforementioned procedure of spectra subtraction also self-suppresses this spectral contribution.

## 4. Conclusion

Fig. 6 displays a general picture of specific electron transitions coupled with the basic process of the threshold core level excitation in any adsorbed system.

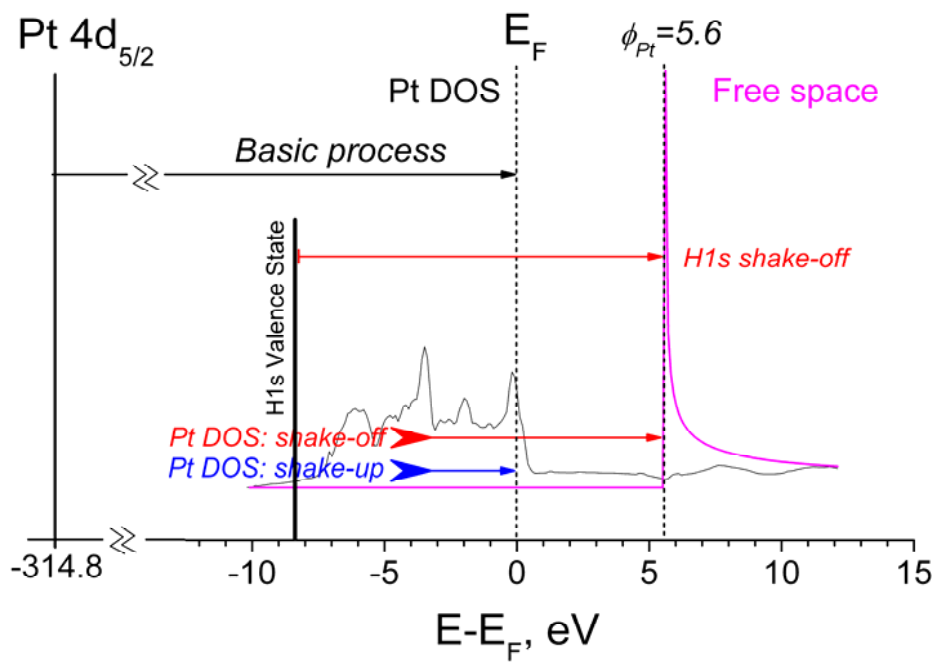

**Fig. 6**  The general picture of electron transitions, as an example, in the H/Pt(100) adsorbed system exposed to the primary electron beam of above E(Pt4d$_{5/2}$) core level energy.

These transitions are independent channels of the primary electron energy consumption and can be classified as follows.

*(1)* A sequential shake-off electron transition from the valence states of the adsorbed species to vacuum level (indicated in Fig. 6 with a red arrow related to H1s state). The probability of this process is represented by the second term of relation (4). According to energy conservation the spectral feature is then observed at the incident electron energy $E_p = E(Pt_{4d_{5/2}}) + \varepsilon_V + \varphi$, where $E(Pt_{4d_{5/2}})$ is a discrete core level energy; $\varphi$ is a work function; $\varepsilon_V$ is a valence state energy of the adsorbed species. Therefore this channel provides a direct experimental data on the density of valence states $\sigma_V(\varepsilon)$ of the adsorbed layer as a single or a set of CEE satellites above $E_F$ as takes place in Figs. 2, 3.

*(2)* In the same terms the probability of shake-off electron transition from the occupied part of Pt 5d-band to vacuum level is also represented by the second part of relation (4) (indicated in Fig. 6 with a red arrow related to Pt DOS). The corresponding spectral satellites then appear in the energy range of $|E(PtDOS)| + \varphi$ and reveal the Pt DOS structure relative the reference point $E_F + \varphi$ as in Figs. 2, 3, 5. Note that electron transitions *(1)* and *(2)* revealed by DAPS are similar to those in a well-proven UPS technique. The only difference concerns the energy source – an overexcited core electron and photon, respectively.

*(3)* The probability of shake-up electron transition from the Pt valence band to vacant state at $E_F$ is represented by the first part of relation (4) and indicated in Fig. 6 with a blue arrow. Similarly to item *(2)* the spectral structure reproduces the Pt DOS right above $E_F$, but without the work function shift, as takes place in Figs. 4, 5.

Fig. 6 displays the transitions between one-electron states and should be supplemented with the collective electron shake-up transitions related to multiple plasmon excitations, which can be also regarded as another kind of the conjugate electron excitation.

To recapitulate, inelastic electron scattering by the adsorbed system is characterized by two peculiar channels related to specific coupled electron excitations. The first channel provides a direct experimental information on the valence state structure of the adsorbed species as well as on the substrate DOS. This kind of excitation points to such a strong integration of the substrate and adsorbate DOS as in a single molecule. The second channel is responsible for multiple plasmon oscillations. The respective satellites in extended DAPS spectra attributing both channels accompany the threshold excitation of both $Pt4d_{5/2}$ and $Pt4d_{3/2}$ core levels. CEE effect is supposed to be a general regularity of electron-solid interaction and a useful tool for fingerprinting the adsorbed layer at the molecular level.